\documentclass[a4paper,12pt]{article}
\pdfoutput=1

\usepackage{
graphicx,
hyperref,
amsmath,
amssymb,
charter,
xcolor,
ifluatex,
cite,
longtable,
booktabs,
multirow,
bbold}
        
\usepackage{colortbl}

\definecolor{Gray}{gray}{0.95}
\definecolor{RGray}{gray}{0.85}
\definecolor{CGray}{gray}{0.92}

\usepackage{multicol}
\definecolor{tit}{rgb}{0.1,0.2,0.4}
\definecolor{blus}{cmyk}{1,1,0,0.6}
\definecolor{verde}{cmyk}{0.92,0,0.59,0.25}
        
\newcommand{\no}{\nonumber\\}

\usepackage{devanagari}
\usepackage{tipa}
\usepackage{amsmath,amssymb,amsfonts, bm}
\usepackage{dsfont}
\usepackage{epsfig}
\usepackage{graphicx}
\usepackage{slashed}
\usepackage{color}

\usepackage{caption}
\usepackage{subcaption}
\usepackage{slashed}

\usepackage{commath}
\usepackage{calc}

\newcommand{\be}{\begin{equation}}
\newcommand{\ee}{\end{equation}}

\newcommand{\bea}{\begin{eqnarray}}
\newcommand{\eea}{\end{eqnarray}}

\newcommand{\bfig}{\begin{figure}}
\newcommand{\efig}{\end{figure}}

\usepackage{commath}
\usepackage{calc}

\makeatletter
\newcommand*{\rom}[1]{\expandafter\@slowromancap\romannumeral #1@}
\makeatother

\begin{document}
\allowdisplaybreaks
\vspace*{-2.5cm}
\begin{flushright}
{\small
IIT-BHU
}
\end{flushright}

\vspace{2cm}

\begin{center}
{\LARGE \bf \color{tit}
A low scale left-right symmetric mirror model}\\[1cm]

{\large\bf Gauhar Abbas$^{a,b}$\footnote{email: gauhar.phy@iitbhu.ac.in}}  
\\[7mm]
{\it $^a$ } {\em Department of Physics, Indian Institute of Technology (BHU), Varanasi 221005, India}\\[3mm]
{\it $^b$ } {\em Theoretical Physics
Division, Physical Research Laboratory, Navrangpura, Ahmedabad 380
009, India}\\[3mm]

\vspace{1cm}
{\large\bf\color{blus} Abstract}
\begin{quote}
A left-right symmetric mirror model restoring parity at a high scale in a way such that the mirror fermions and mirror gauge sector simultaneously could exist at TeV scale is discussed.  We also provide an  ultraviolet completion of the model with vector-like fermions, and discuss some theoretical and phenomenological implications of this model.
\end{quote}

\thispagestyle{empty}
\end{center}

\begin{quote}
{\large\noindent\color{blus} 
}

\end{quote}

\newpage
\setcounter{footnote}{0}

\section{Introduction}
There are two classes of parity restoring models in general.  In the first class, the  right-handed degrees of freedom of the standard model (SM) and three right-handed neutrinos are accommodated in the fundamental representation of the gauge group $SU(2)_R$ \cite{Pati:1974yy,Mohapatra:1974hk,Senjanovic:1975rk,Mohapatra:1979ia}.    The second class of models have mirror fermions in the fundamental representation of the gauge group  $SU(2)_R$ instead of the right-handed degrees of freedom of the SM \cite{Abbas:2016xgj,Abbas:2016qqc,Babu:1989rb,Foot:1991bp,Silagadze:1995tr,Berezhiani:1995yi,Gu:2012in,Chakdar:2013tca,Cui:2011wk,Cui:2012mq,Gu:2014mga,Gu:2017mkm,Lavoura:1997pq,Abbas:2017vle}.  In these models, the right-handed degrees of freedom of the SM and three right-handed neutrinos are treated as singlets under the gauge group $SU(2)_R$. The latest phenomenological status of these models can be found in Refs.\cite{Gu:2012in,Chakdar:2013tca,Cui:2011wk,Cui:2012mq,Lindner:2016lpp,Patra:2015bga,Bertolini:2014sua}.

It is quite disappointing that mirror gauge sector of the models having mirror fermions and mirror symmetries turns out to be extremely heavy.  The reason lies in the fact that parity invariance makes the Yukawa couplings of the SM and mirror sector identical.  For instance, the Yukawa Lagrangian in Ref. \cite{Gu:2012in} with gauge symmetry $SU(2)_L \times SU(2)_R \times U(1)_Y$ is 
\be
{\mathcal{L}}_{Y} = \Gamma \left(  \bar{\psi_L}  \varphi_L  \psi_R +   \bar{\psi^{\prime}_{R}}  \varphi_R \psi^{\prime}_{L} \right)+ {\rm H.c.},
\ee
where $\psi^\prime$ fermions are mirror counter-parts of the SM $\psi$ fermions.  They are singlet under the SM gauge group $SU(2)_L$ and charged under the mirror gauge group $SU(2)_R$.  The scalar  Higgs field $\varphi_L$ is doublet under the SM gauge group $SU(2)_L$, and singlet under the mirror gauge group $SU(2)_R$.  Similarly, scalar  Higgs field $\varphi_R$ is doublet under the gauge group $SU(2)_R$, and singlet under the SM gauge group $SU(2)_L$.  $\Gamma$ is $3 \times 3$ matrix in family space.

We have not seen these mirror fermions at the Large Hadron Collider (LHC) around TeV scale yet.   The mass of the lightest charged lepton is given as  $m_{e^\prime} = m_e  \langle \varphi_R \rangle /  \langle \varphi_L \rangle$ where $m_e$ is mass of the electron, $\langle \varphi_L \rangle=246$ GeV is the vacuum expectation value (VEV) of the SM Higgs field, and $\langle \varphi_R \rangle$ is VEV of the Higgs field $\varphi_R$. Now, for instance, $m_e = 0.511$ MeV and  $\langle \varphi_R \rangle = 5 \times 10^8$ GeV, the mass of the lightest charged lepton is $m_e^\prime = 1038.65$ GeV that could be looked for at the LHC.    

Hence, for sufficiently heavy mirror fermions to search at the  LHC, we need a large parity breaking scale around $10^8$ GeV.  Hence, mirror gauge bosons corresponding to the gauge group $SU(2)_R$ have masses of order $10^8$ GeV \cite{Gu:2012in,Chakdar:2013tca}.  Moreover, requirement of small neutrino masses further increases scale of parity breaking.  Furthermore, these mirror gauge bosons are out of the reach of the LHC, and being so heavy may not be able to search  in near future.

In this paper, we propose a low scale  left-right symmetric mirror model  which restore parity  in a way such that mirror gauge sector and mirror fermions  can exist simultaneously at TeV scale.   An ultraviolet (UV) completion of this model with vector-like fermions is also discussed.  

We organize this paper as follows: In section \ref{sec2}, we discuss our model.  Section \ref{sec3} has some theoretocal and phenomenological implications of the model which includes strong $CP$ problem, dark matter and collider signatures. An ultra-violet completion of the model is presented in section \ref{sec4}.   We conclude in section \ref{sec5}.

\section{Low scale left-right symmetric mirror model}
\label{sec2}
Our model can be described by the following field transformations under $SU(3) \times SU(2)_L \times SU(2)_R \times U(1)_{Y^\prime}$:
\begin{eqnarray}
l_{L}={\begin{pmatrix} \nu \\ e \end{pmatrix}}_L\sim (1,2,1,-1) &,& e_R \sim (1,1,1,-2);\nonumber\\
q_{L} ={\begin{pmatrix} u \\ d \end{pmatrix}}_L\sim (3,2,1,\frac{1}{3}) &,& u_R \sim (3,1,1,\frac{4}{3})~~~,~~ d_R \sim(3,1,1,-\frac{2}{3});\nonumber\\
l_{R}^{\prime}={\begin{pmatrix} \nu^\prime \\ e^\prime \end{pmatrix}}_R\sim (1,1,2,-1) &,& e^\prime_L \sim (1,1,1,-2);\nonumber\\ 
q_{R}^\prime ={\begin{pmatrix} u^\prime \\ d^\prime \end{pmatrix}}_R\sim (3,1,2,\frac{1}{3}) &,& u^\prime_L \sim (3,1,1,\frac{4}{3})~~~,~~ d_L^\prime \sim(1,1,1,-\frac{2}{3});
\label{eq1} 
\end{eqnarray}
where $l_L$, $q_L$ are the SM doublets of leptons and quarks, and $e_R$, $\nu_{eR}$, $d_R$ and $u_R$ are the SM singlets. $l_R^\prime$, $q_R^\prime$, $e_L^\prime$, $\nu_{eL}^\prime$, $d_L^\prime$ and $u_L^\prime$ denote mirror fermions and their quantum numbers. The field transformations for second and third families are identical to the first family.  

The fermionic and gauge fields under  parity transform as:
\begin{eqnarray}
&&\psi_L \longleftrightarrow \psi^{\prime}_R,~ \psi_R \longleftrightarrow \psi^{\prime}_L, ~
\mathcal{W}_L \longleftrightarrow  \mathcal{W}_R,~
  \mathcal{B}_\mu \longleftrightarrow  \mathcal{B}_\mu,~\mathcal{G}_{\mu \nu} \longleftrightarrow  \mathcal{G}_{\mu \nu},
\end{eqnarray}
where $\psi_L$ is a doublet  of the  gauge groups $SU(2)_L$, and  $\psi^{\prime}_R$ is a doublet of the  gauge group  $SU(2)_R$.   $\psi_R$ and $\psi^{\prime}_L$ are singlets under either of them.  $ \mathcal{W}_L $ is the gauge field corresponding to the gauge group $SU(2)_L$, and $ \mathcal{W}_R $ is the gauge field of the gauge symmetry $SU(2)_R$. $\mathcal{B}_\mu $ represents the gauge field corresponding to the gauge symmetry $U(1)_Y$.  $\mathcal{G}_{\mu \nu}$ denotes gluon field strength tensor. 
 
The spontaneous symmetry breaking (SSB) occurs in the following way:
 {
\begin{equation}
SU(2)_L\times SU(2)_R \times U(1)_{Y^{\prime}} \to SU(2)_L\times U(1)_{Y} \to U(1)_{EM}.
\end{equation}} 

The above symmetry breaking pattern is achieved by introducing two Higgs doublets which transform in the following way under $SU(3)_c \times SU(2)_L \times SU(2)_R  \times U(1)_{Y^\prime} $:
\begin{eqnarray}
&&\varphi_L = {\begin{pmatrix} \varphi^+ \\ \varphi^0 \end{pmatrix}}_L\sim(1,2,1,1),~\varphi_R= {\begin{pmatrix} \varphi^+ \\ \varphi^0 \end{pmatrix}}_R\sim(1,1,2,1),
 \end{eqnarray} 
and behave under parity as follows:
\begin{eqnarray}
&&\varphi_L \longleftrightarrow \varphi_R
\end{eqnarray} 
Besides doublets, two real gauge scalar singlet fields $\chi$ and $\chi^\prime$ are also needed to provide masses to fermions as discussed later.  Singlets scalar fields have following quantum numbers under $SU(3)_c \times SU(2)_L \times SU(2)_R  \times U(1)_{Y^\prime} $:
\begin{eqnarray}
 \chi:(1,1,1,0),~ \chi^\prime:(1,1,1,0),
 \end{eqnarray} 
and their behaviour under parity is described as,
\begin{eqnarray}
\chi \longleftrightarrow \chi^{\prime}.
\end{eqnarray}

\begin{table}[h]
\begin{center}
\begin{tabular}{|c|c|c|c|}
  \hline
  Fields             &        $\mathcal{Z}_2$                    & $\mathcal{Z}_2^\prime$       &        $\mathcal{Z}_3$     \\
  \hline
  $\psi_R$                 &   +  &     -      &    $\omega$                          \\
  $\chi$                        & +  &      -     &      $\omega^2$                                         \\
  $ \psi_L^\prime$     & -  &   +          &   $\omega$                                  \\
  $\chi^\prime$           & - &      +       &    $\omega^2$   \\
  \hline
     \end{tabular}
\end{center}
\caption{The charges of fermionic and singlet scalar fields under $\mathcal{Z}_2$, $\mathcal{Z}_2^\prime$ and $\mathcal{Z}_3$ symmetries where $\omega$ is the cube root of unity.}
 \label{tab1}
\end{table}    

For having mirror gauge sector and mirror fermions simultaneously at TeV scale, we note that fermion-scalar interactions are not governed by any symmetry in the SM and put into the SM by hand.  Their description through the Yukawa operator is a selection.  However, it is possible that these interactions are actually described by dimensional-5 operator.

Hence, for this purpose, a pair of discrete symmetries  $\mathcal{Z}_2$ and  $\mathcal{Z}_2^\prime$ is imposed on the fermionic fields $\psi_R$, $\psi_L^\prime$ and scalar singlets $\chi$, $\chi^\prime$, keeping all other fields even under $\mathcal{Z}_2$ and  $\mathcal{Z}_2^\prime$.  This is shown in Table \ref{tab1}.   Moreover, to protect the stability of the scalar potential from the hierarchical vacuum-expectation-values (VEVs) of the singlet scalar fields  $\chi$ and $\chi^\prime$, we impose an additional discrete symmetry  $\mathcal{Z}_3$ on the right-handed fermions and singlet scalar fields.

\subsection{Fermion-scalar sector of the model}
The discrete symmetries $\mathcal{Z}_2$ and  $\mathcal{Z}_2^\prime$ forbid the Yukawa operator, and mass term, for instance,  for first family leptons is given by dimensional-5 operator,
\bea
\label{mass1}
{\mathcal{L}}_{mass} &=& \dfrac{1}{\Lambda} \left[ \bar{l_L} \left( \Gamma_1   \varphi_L \chi + \Gamma_2 \tilde{\varphi}_L  \chi  \right) e_R + \bar{l^{\prime}_{R}}  \left( \Gamma_1^{\prime }  \varphi_R \chi^\prime  + \Gamma_2^{\prime} \tilde{\varphi}_R  \chi^\prime  \right) e^{\prime}_{L} \right] \\ \nonumber
&+& \dfrac{1}{\Lambda} \left[  \rho_1 ~ \bar{l_L}    \varphi_L \varphi_R^{ \dagger}  l_R^\prime +  \rho_2 ~ \bar{l_L}    \tilde{\varphi}_L \tilde{\varphi}_R^{ \dagger}  l_R^\prime   \right]+ {\rm H.c.},
\eea
where $\Gamma_i = \Gamma_i^{\prime}$ ($i=1,2$) due to parity and, $\Gamma_i$, $\rho$, $ \sigma$ are $3 \times 3$ matrices in family space.  $ \tilde{\varphi} = i \sigma_2 \varphi^*$ is charge conjugated Higgs field, and $\sigma_2$ is the second Pauli matrix.  We can write a similar Lagrangian for other fermions.

\begin{figure}[t!]
 \begin{center}
  \includegraphics[scale=0.7]{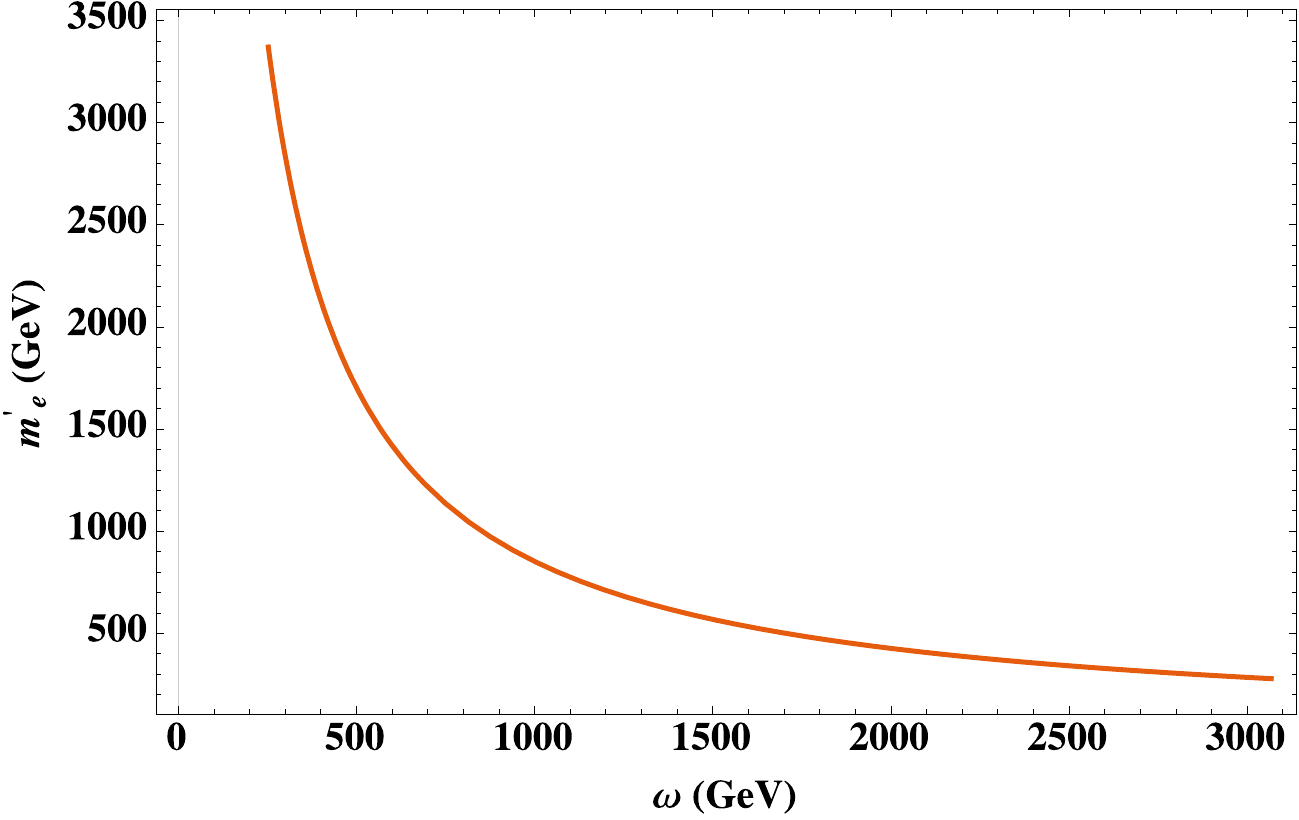}
    \caption{A bound on the mass of the lightest mirror charged lepton assuming   $  \langle \chi^\prime \rangle =  10^8$ GeV, and  $\langle \chi \rangle  = \omega $.}
    \label{scale_mirror}
    \end{center}
\end{figure}
We need the pattern of the SSB such that $ \langle \chi^\prime \rangle >>   \langle \varphi_R \rangle  >>    \langle \varphi_L \rangle$ and $ \langle \chi^\prime \rangle >> \langle \chi \rangle $.  This will result in a mirror gauge and fermionic sector at TeV scale simultaneously.   Now, mass of the mirror counter-part of the electron is 
$m_{e^\prime} \approx m_e  \langle \varphi_R \rangle  \langle \chi^\prime \rangle  / \langle \varphi_L \rangle \langle \chi \rangle$.

The LHC has searched for mirror fermions, and has excluded them upto $690$GeV\cite{Aad:2015tba}.  However, these searches are model dependent.  The CMS collaboration has searched for an extra $W$ boson and has excluded it  upto 4.1 TeV\cite{Khachatryan:2016jww}.   Hence, being conservative, we can assume $\langle \varphi_R \rangle \geq 4.1$ TeV.  The SM extended by a real singlet scalar field is studied in Ref. \cite{Robens:2016xkb}, and range of  $\langle \chi \rangle  $ is given between 2.5 GeV to 3075 GeV.  Using these numbers, we provide a rough bound between approximately 500 GeV to 3400 GeV on the mass of the lightest charged mirror lepton in Fig.\ref{scale_mirror} where we have chosen  $  \langle \chi^\prime \rangle =  10^8$ GeV.  However, this bound depends on the chosen value of  $  \langle \chi^\prime \rangle$.   Thus, we observe that remarkably new mirror gauge  and fermionic sector could exist simultaneously around TeV scale.  

The Majorana mass term for neutrinos is provided by the following equation:
\be
\label{mass2}
{\mathcal{L}}^{\nu}_{Majorana} = \dfrac{c}{\Lambda} \left[ \bar{l_{L}^c}    \tilde{\varphi}_{L}^* \tilde{\varphi}_L^\dagger  l_L  +  \bar{l_{R}^{\prime c}}    \tilde{\varphi}_{R}^* \tilde{\varphi}_R^\dagger  l_R^\prime  \right]    + {\rm H.c.},
\ee
and masses of neutrinos can be recovered through seesaw mechanism.

Now we discuss masses of fermions.  They are given by Eq.(\ref{mass1}) and (\ref{mass2}).  For instance, the Lagrangian for the down type quark and its mirror counter-part can be written as,
\begin{eqnarray}
\label{Lmass}
{\cal L}_d &=& \dfrac{\Gamma_d}{\Lambda} \left(\bar q_L \varphi_L d_R \chi  + \bar { q}^\prime_R  \varphi_R d^\prime_L \chi^\prime \right) +\dfrac{ \rho_d}{\Lambda}  ~ \bar{q_L}    \varphi_L \varphi_R^{ \dagger}  q_R^\prime   + {\rm H.c.}\nonumber\\
&=& {\begin{pmatrix} \bar d_L & \bar { d}^\prime_L \end{pmatrix}}{\begin{pmatrix}\frac{\Gamma_d v_L \omega}{2 \Lambda} & \frac{ \rho_d v_L v_R}{2 \Lambda} \\ 0  & \dfrac{\Gamma_d^* v_R \omega^\prime}{2 \Lambda}\end{pmatrix}}{ {\begin{pmatrix} d_R \\  { d}^\prime_R \end{pmatrix}}}~+~{\rm H.c.},
\end{eqnarray} 
where the mass matrix in general is $6 \times 6$.

The mass matrices can be  diagonalized through bi-unitary transformations given as,
\be
\label{bidiag}
\left( \begin{array}{c} u \\ u^\prime \end{array} \right)_{L,R}
=
\left( \begin{array}{c} X_{u} \\ Y_{u} \end{array} \right)_{L,R}
\left( \begin{array}{c} u \\ u^\prime \end{array} \right)_{L,R} \ \mbox{\rm and} \
\left( \begin{array}{c} d \\ d^\prime \end{array} \right)_{L,R}
=
\left( \begin{array}{c} X_{d} \\ Y_{d} \end{array} \right)_{L,R}
\left( \begin{array}{c} d \\ d^\prime \end{array} \right)_{L,R},
\ee
where $X_{u,d}$ and $Y_{u,d}$ are $3 \times 6$, and CKM matrices are given by $V_{CKM} = X_{uL}^\dagger X_{dL}$ and $V_{CKM}^\prime = Y_{uR}^\dagger Y_{dR}$.

\subsection{Gauge-scalar sector of the model}

The gauge-scalar interactions of the model are described by  the following Lagrangian:
\begin{eqnarray}
{\cal L}_{GS}& =& \left({\cal D}_{\mu,L}\varphi_L \right)^\dagger\left({\cal D}^{\mu}_L \varphi_L \right)
+  \left({\cal D}_{\mu,R}\varphi_R \right)^\dagger\left({\cal D}^{\mu}_R \varphi_R \right) ,
\label{ktl}
\end{eqnarray}
where, ${\cal D}_{L,R}$ are the covariant derivatives given by,
\begin{eqnarray}
\mathcal{D}_{\mu,L} (\mathcal{D}_{\mu,R}) &=& \partial_\mu + i g  \dfrac{\tau_a}{2} \mathcal{W}^a_{\mu,L} (\mathcal{W}^{a}_{\mu,R}) +ig^\prime \frac{Y^\prime}{2} B_\mu,
\end{eqnarray}
where, $\tau_a$'s are the Pauli matrices.  $g$ is the common coupling of the gauge groups $SU(2)_L $ and $ SU(2)_R$.  The coupling constant  $g^\prime$ corresponds to the gauge group $U(1)_{Y^\prime}$.
 
The  charged gauge bosons masses after the SSB are given by,
\begin{equation}
M_{W^\pm_{L}}~=~\frac{1}{2}g v_L,~~M_{ W^{ \pm}_{R}}~=~\frac{1}{2} g v_R.
\end{equation}  

The neutral gauge boson mass matrix in the basis ($W_L^3,~ W_R^3,~B$) is given by,
\begin{equation}
M=\frac{1}{4}{\begin{pmatrix} g^2v_L^2 & 0 & -gg^\prime v_L^2 \\ 0 & g^2  v_R^2 & -gg^\prime  v_R^2 \\ -gg^\prime v_L^2 &  -gg^\prime  v_R^2 & g^{\prime 2}(v_L^2+ v_R^2) \end{pmatrix}}.
\label{MZ}
\end{equation}

The matrix in Eq.(\ref{MZ}) can be diagonalized through an orthogonal transformation $\mathcal{T}$ which transforms the weak eigenstates: ($W_L^3, W_R^3,~B$) into  the physical mass eigenstates: ($Z_L,~Z_R,~\gamma$);
\begin{equation}
{\begin{pmatrix} W_L^3 \\  W_R^3 \\ B \end{pmatrix}}=\mathcal{T} {\begin{pmatrix} Z_L \\  Z_R \\ \gamma \end{pmatrix}}.
\label{trans}
\end{equation}

The physical masses of the neutral gauge bosons are then written as,
\begin{equation}
M_{Z_L}^2= \frac{1}{4}v_L^2g^2\frac{g^2+2g^{\prime 2}}{g^2+g^{\prime 2}}\left[1-\frac{g^{\prime 4}}{\left(g^2+g^{\prime 2}\right)^{ {2}}}\epsilon\right],
M_{ Z_R}^2= \frac{1}{4} v_R^2\left(g^2+g^{\prime 2}\right)\left[1+\frac{g^{\prime 4}}{\left(g^2+g^{\prime 2}\right)^{ {2}}}\epsilon\right],
\label{MZ2}
\end{equation}
where  $\epsilon = v_L^2/ v_R^2$.  Since $v_R >> v_L$, we have ignored terms of order ${\cal O}(\epsilon^2)$  in Eq.(\ref{MZ2}).

The orthogonal matrix $\mathcal{T}$ in Eq.(\ref{trans}) can be parametrized in terms of mixing angle $\theta_{W}$ which is given by the following equation:
\begin{eqnarray}
{\rm cos}^2\theta_{W} = \left(\frac{M_{W_L}^2}{M_{Z_L}^2}\right)_{\epsilon=0}=\frac{g^2+g^{\prime2}}{g^2+2g^{\prime 2}}.
\end{eqnarray}
and transformation matrix $\mathcal{T}$ can be written as,
\begin{equation}
\mathcal{T} = {\begin{pmatrix}
-{\rm cos}\theta_{W} & -  \dfrac{{\sqrt{\rm cos 2 \theta_{W}}} {\rm tan}^2 \theta_{W}}{{\rm cos} \theta_{W}} \epsilon & {\rm sin} \theta_{W} \\
{\rm sin} \theta_{W} {\rm tan} \theta_{W} \left[ 1+\frac{{\rm cos} 2 \theta_{W}}{{\rm cos}^4\theta_{W}} \epsilon\right] & -\dfrac{{\sqrt{\rm cos 2 \theta_{W}}}}{{\rm cos} \theta_{W} } \left[1-{\rm tan}^4 \theta_{W} \epsilon \right] & {\rm sin}\theta_{W} \\
{\rm sin}\theta_{W} \dfrac{{\sqrt{\rm cos 2 \theta_{W}}}}{{\rm cos} \theta_{W} } \left[ 1-\frac{{\rm tan}^2 \theta_{W}}{{\rm cos}\theta_{W}} \epsilon\right] & {\rm tan} \theta_{W} \left[1+{\rm tan}^2 \theta_{W} \frac{{\rm cos} 2 \theta_{W}}{{\rm cos}^2\theta_{W}} \epsilon \right] &  \sqrt{\rm cos 2 \theta_{W}}
\end{pmatrix}}.
\end{equation}
We note that  the third column of that matrix is unchanged by further terms of order $\epsilon^2$ in the transformation matrix $\mathcal{T}$.

The couplings of the original symmetries and the residual symmetry are related by the following equation:
\begin{equation}
g=\frac{e}{{\rm sin}\theta_{W}},~~g^\prime=\frac{e}{\sqrt{\rm cos 2 \theta_{W}}},~~
\frac{1}{e^2}=\frac{2}{g^2}+\frac{1}{g^{\prime 2}}.
\end{equation}

Now, we can define the following hermitian idempotent  matrices:
\be
H_u = X_{u}^\dagger X_{u} \ \mbox{\rm and} \
H_d = X_{d}^\dagger X_{d},
\ee
where dimensions of the matrices $H_{u,d}$ are $6  \times 6$.  

The neutral current Lagrangian can be written as,
\bea
 & &
\overline{u} \gamma_\mu
\left[
- \frac{2}{3} e A^\mu
+ \frac{g}{\cos \theta_W} Z^\mu_L
\left( \frac{1}{2} H_u \gamma_L - \frac{2}{3} \sin^2 \theta_W \right)
\right.
\no
 & &
\left.
+ \frac{g \sqrt{\cos^2 \theta_W - \sin^2 \theta_W}}{\cos \theta_W} Z^{ \mu}_R
\left( - \frac{2}{3} \frac{\sin^2 \theta_W}{\cos^2 \theta_W - \sin^2 \theta_W}
+ \frac{1}{2} \frac{\cos^2 \theta_W}{\cos^2 \theta_W - \sin^2 \theta_W} \gamma_R
- \frac{1}{2} H_u \gamma_R \right)
\right]
u
\no
 & &
+ \overline{d} \gamma_\mu
\left[
\frac{1}{3} e A^\mu
+ \frac{g}{\cos \theta_W} Z^\mu_L
\left( \frac{1}{3} \sin^2 \theta_W - \frac{1}{2} H_d \gamma_L \right)
\right.
\no
 & &
\left.
+ \frac{g \sqrt{\cos^2 \theta_W - \sin^2 \theta_W}}{\cos \theta_W} Z^{ \mu}_R
\left( \frac{1}{3} \frac{\sin^2 \theta_W}{\cos^2 \theta_W - \sin^2 \theta_W}
- \frac{1}{2} \frac{\cos^2 \theta_W}{\cos^2 \theta_W - \sin^2 \theta_W} \gamma_R
+ \frac{1}{2} H_d \gamma_R \right)
\right]
d, \quad \quad\quad
\eea 
where $\gamma_L = (1 - \gamma_5) / 2$ and  $\gamma_R = (1 + \gamma_5) / 2$

\subsection{Scalar potential of the model}
The most general scalar potential of the model reads,
\begin{eqnarray}
V &=& -\mu_L^2  \varphi_L^\dagger \varphi_L    -\mu_R^2  \varphi_R^{ \dagger} \varphi_R   - \mu_\chi^2  \chi^2   - \mu_{\chi^\prime }^{2}  \chi^{\prime 2}  
+  \lambda_1 \Bigl(   (\varphi_L^\dagger \varphi_L)^2 +   (\varphi_R^{\dagger} \varphi_R)^2  \Bigr)  \\ \nonumber
&+& \lambda_2  \varphi_L^\dagger \varphi_L \varphi_R^\dagger \varphi_R +  \rho     \Bigl(   \varphi_L^\dagger \varphi_L \chi +   \varphi_R^{\dagger} \varphi_R \chi^\prime \Bigr)  . 
\end{eqnarray}
This potential is parity invariant except  for mass terms of scalar fields which constitutes a soft parity breaking. This is required  to break parity spontaneously in the gauge sector of the model such that  $ \langle \chi^\prime \rangle = \omega^\prime/\sqrt{2} >> \langle \varphi_R \rangle=v_R/\sqrt{2}  >>    \langle \varphi_L \rangle=v_L/\sqrt{2}$ and $ \langle \chi^\prime \rangle >> \langle \chi \rangle=\omega/\sqrt{2} $.    All couplings of the scalar potential are real, and VEVs of the scalar doublets can be made real through the gauge symmetry.   Moreover, we have added soft symmetry breaking terms in the potential.

\section{Strong $CP$ problem, dark matter and phenomenological signatures}
\label{sec3}
We note that parity is only softly broken symmetry in our model.  Therefore, mass matrices, for instance given in Eq.\ref{Lmass}, are real.  Hence, strong $CP$ phase is zero at tree-level. At one-loop level, diagrams having neutral scalars or neutral massive gauge bosons in the loop may generate complex phase in the mass term of the quarks.  The traces of the one-loop amplitude of these diagrams depend on  the product matrices $H_u^\dagger H_u$ and  $H_d^\dagger H_d$.  The off diagonal elements of  $H_u^\dagger H_u$ and $H_d^\dagger H_d$ are complex.  However, daigonal mass terms are real.  Hence, one-loop contribution to the strong $CP$ phase is zero\cite{Babu:1989rb}.   At two-loop, there are terms in the traces of two-loop amplitudes which depend on solely on $H_u$  and $H_d$ which already have complex off-diagonal mass terms.  Therefore at two-loops, the strong  $CP$ phase is non-zero.  However, this non-zero contribution will be suppressed by a factor of $ \Bigl( \dfrac{ \langle \varphi_L \rangle \langle  \chi \rangle }{ \langle \varphi_R \rangle \langle  \chi^\prime \rangle} \Bigr)^2$\cite{Babu:1989rb}.  Since $ \langle  \chi^\prime \rangle $ is expected to be very large, this non-zero contribution to the strong phase is extremely small.  Hence, the model provide a solution of the strong $CP$ problem.

This model may provide multicomponent dark matter.  It may be achieved if the mass term $\dfrac{ \rho_d}{\Lambda}  ~ \bar{q_L}    \varphi_L \varphi_R^{ \dagger}  q_R^\prime  $ in Eq.\ref{Lmass} is highly suppressed.  This suppression may emerge if the renormalization scale $\Lambda$ of our model is very large (for instance, GUT scale) such that $\Lambda >> M_{\psi^\prime}$ where $M_{\psi^\prime}$ denotes the scale of the mirror fermionic sector.  In this case, mirror sector is  practically decoupled from the ordinarry sector, and coupled to the ordinary sector only through the pure scalar sector.  Hence, mirror fermionic and gauge sector may act like dark matter which is already investigated in literature\cite{Foot:2014mia}. 

In this effective decoupling of the mirror-sector, the two-loop contribution to the strog $CP$ phase is also effectively zero, and model is equivalent to the model in Ref.\cite{Barr:1991qx} where non-zero contribution to the strong $CP$ phase appears at three-loops, and is extremely small.

\begin{figure}[t!]
 \begin{center}
  \includegraphics[scale=0.5]{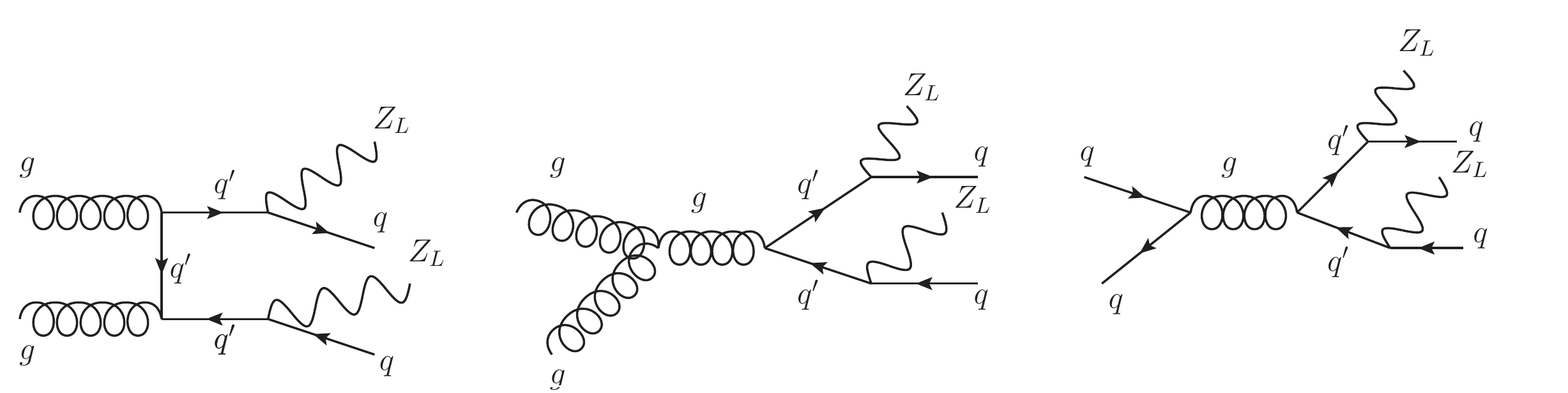}
    \caption{The pair production of the mirror quarks at the LHC and their subsequent decay to the SM $Z_L$ boson and a quark.}
    \label{fig1}
     \end{center}
\end{figure}
Phenomenologically interesting scenario is where $\Lambda$ is not very large, and within the reach of the LHC.  In this case, there are flavour changing neutral current interactions at tree level in the model due to the mixing of mirror and ordinary fermions as well as due to vector-like fermions(to be discussed later).  However, these interactions are suppressed by $v_L^2/v_R^2$  or by $v_L^2/M^2$ where $M$ is the mass of vector-like fermion\cite{Branco:1986my,Lavoura:1992qd,Lavoura:1992np}.  Hence, low energy phenomenological consequences will appear, for instance, in flavour changing neutral current processes which are highly suppressed in the SM.  For example, apart from the SM box diagrams, new diagrams, having $W_R$ in the box, contributing to the  $\Delta F = 2$ transitions,  of $K$ and $B$ mesons may place non-trivial constraints on the masses of the new gauge bosons  and mirror fermions.

We observe that due to mixing of the SM and mirror fermions, the mirror fermions can decay into a SM $W_L$ or $Z_L$ boson in association with a SM fermion. Since mirror quarks can couple to gluons,  for illustration, we show the pair production of the mirror quarks in Fig.\ref{fig1} at the LHC via gluon-gluon and quark-antiquark initial states.
   

\section{Ultra-violet completion of the model}
\label{sec4}
Now we present a UV completion of the model discussed in this paper.  For this purpose, we add  one vector-like isosinglet up type quark, one vector-like isosinglet down type quark, and one vector-like  iso-singlet charged leptons.  They transform under $SU(3)_c \times SU(2)_L \times SU(2)_R  \times U(1)_{Y^\prime} $ as follows:
\begin{eqnarray}
 Q &=& U_{L,R}:(3,1,1,\dfrac{4}{3}); D_{L,R}:(3,1,1,-\dfrac{2}{3}); 
  L=  E_{L,R}:(1,1,1,-2).
\end{eqnarray}

  \begin{figure}[t!]
 \begin{center}
  \includegraphics[scale=0.7]{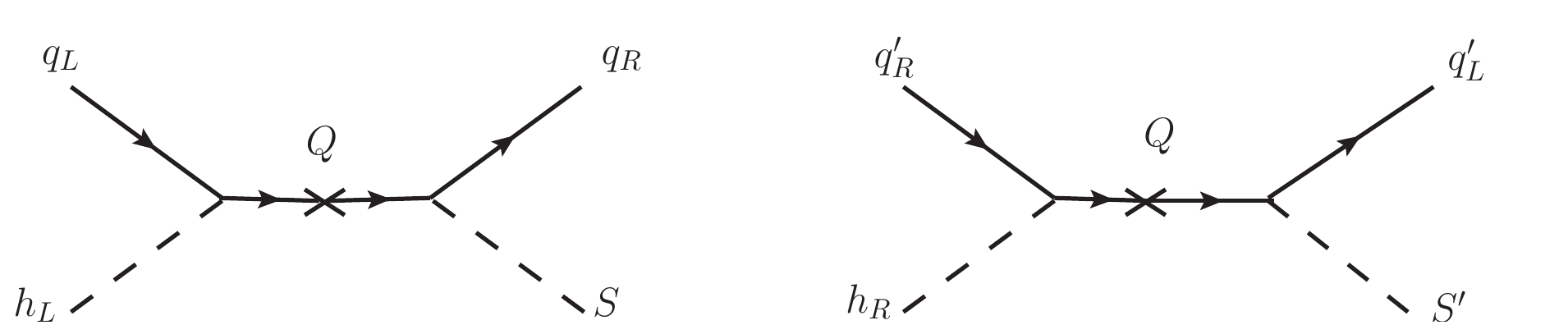}
    \caption{The mass term of the SM and mirror fermions in Eq.(\ref{mass1}) after introducing vector-like iso-singlet fermions.  Here, $q_L$, $q_R$  and $h_L$ are the SM quarks and Higgs boson, $q_R^\prime$, $q_L^\prime$  and $h_R$ correspond to mirror sector,  $Q$ is a vector-like iso-singlet quark, $S$ and $S^\prime$ are scalar particles corresponding to the scalar field $\chi$ and $\chi^\prime$.}
    \label{fig6}
    \end{center}
\end{figure}

The mass term for vector-like fermions can be written as,
\begin{eqnarray}
\mathcal{L}_{V} &=& M_U \bar{U}_L U_R + M_D \bar{D}_L D_R   
+  M_E \bar{E}_L E_R   + {\rm H.c.}.
\end{eqnarray}
The interactions of vector-like fermions with the SM  fermions are given by,
\begin{eqnarray}
\mathcal{L}_{Vff^\prime}^\prime = y^\prime \left[  \bar{q}_L \varphi_L Q_R  + \bar{q}_R^\prime \varphi_R Q_L \right] + c^\prime \left[  \bar{l}_L \varphi_L L_R  + \bar{l}_R^\prime \varphi_R L_L \right] + {\rm H.c}.
\end{eqnarray}

The interactions of singlet SM and mirror fermions with vector-like fermions are given by, 
\begin{eqnarray}
\mathcal{L}_{Vff^\prime}^{\prime \prime} = y^{\prime \prime} \left[  \bar{Q}_L  q_R \chi + \bar{Q}_R q_L^\prime \chi^\prime \right] + c^{\prime \prime} \left[  \bar{L}_L  l_R \chi + \bar{L}_R l_L^\prime \chi^\prime \right] + {\rm H.c}.
\end{eqnarray}
Now the Lagrangian giving masses of the SM and mirror fermions in Eq.(\ref{mass1}) can be realized as shown in Fig.\ref{fig6}.  Vector-like fermions are extensively studied in literature\cite{Lavoura:1997pq,Branco:1986my,Lavoura:1992qd,Lavoura:1992np,Barducci:2013zaa,Barducci:2014ila,Barducci:2014gna,Burgess:1993vc,delAguila:1998tp,AguilarSaavedra:2002kr,Cacciapaglia:2010vn,
Dawson:2012di,Aguilar-Saavedra:2013qpa,Ellis:2014dza,Angelescu:2015kga,Fajfer:2013wca,Alok:2015iha,Alok:2014yua,Chen:2017hak}.

\subsection{Lepton flavour non-universality}
We briefly comment on recently observed anomalies in the flavor-changing neutral current transition $b \rightarrow s l^+ l^-$ and their consequences in the model discussed in this work.    The first deviation from the SM is in the optimised observable $P_5^\prime$ \cite{Descotes-Genon:2013vna} measured by the LHCb of $3.7 \sigma$ significance \cite{Aaij:2013qta}.  Another interesting observable measured by the LHCb which is hinting the lepton flavour universality (LFU)  violation is the ratio $R_K = \mathcal{B}_{B \rightarrow K \mu^+ \mu^-} /  \mathcal{B}_{B \rightarrow K e^+ e^-}$ \cite{Aaij:2014ora}.  More interestingly, the LHCb has recently presented their results on the ratio $R_{K^*} = \mathcal{B}_{B \rightarrow K^* \mu^+ \mu^-} /  \mathcal{B}_{B \rightarrow K^* e^+ e^-}$ showing significant deviation from the SM lepton-flavour universality \cite{Aaij:2017vbb}.    In the model discussed in this paper, LFU violation enters through mixing of the SM and mirror fermions  and  a possible explanation of deviations observed in $P_5^\prime$, $R_{K,K^*}$ may be provided by a contribution due to new heavy vector gauge boson at tree level.

In addition to this, an evidence for an explanation of flavour anomalies  in the quark-level $b \rightarrow s l \bar{l}$ transitions comes from a recent nice paper by Botella et al\cite{Botella:2017caf}.  In this paper, flavour anomalies  in the quark-level $b \rightarrow s l \bar{l}$ transitions are explained by the FCNC effects of vector-like quarks and a heavy neutrino.  The model presented in this work is in fact has naturally vector-like quarks $U, D$ and a heavy neutrino $N$ required in Ref.\cite{Botella:2017caf}.  Hence, an explanation of flavour anomalies  in the quark-level $b \rightarrow s l \bar{l}$ transitions in the model discussed in this paper may be achieved like  in Ref.\cite{Botella:2017caf}.

\section{Conclusion} 
\label{sec5}
Finally we conclude  that the model discussed in this paper restores parity in a way such that mirror gauge and fermionic sectors can coexist at the same scale. There are many theoretical and phenomenological implications of the model such as the strong $CP$ problem, dark matter and  LFU violation along with the different collider signatures.   A detailed phenomenological investigation is under progress.

\end{document}